\documentclass[10pt, conference]{IEEEtran}
\IEEEoverridecommandlockouts
\usepackage{cite}
\usepackage{amsmath,amssymb,amsfonts}
\usepackage{algorithm}
\usepackage{algorithmic}
\usepackage{graphicx}
\usepackage{comment}
\usepackage{textcomp}
\usepackage{subcaption}
\DeclareMathOperator{\Tr}{Tr}
\DeclareMathOperator{\Diag}{diag}
\usepackage{xcolor}
\usepackage{stfloats}
\def\BibTeX{{\rm B\kern-.05em{\sc i\kern-.025em b}\kern-.08em
    T\kern-.1667em\lower.7ex\hbox{E}\kern-.125emX}}
\usepackage{hyperref}

\IEEEoverridecommandlockouts\IEEEpubid{\makebox[\columnwidth]{ 978-1-6654-5975-
4/22~\copyright~2024 IEEE \hfill} \hspace{\columnsep}\makebox[\columnwidth]{ }}

\begin{document}

\title{Fluid Antenna-Assisted Near-Field System\\
}

\author{
\IEEEauthorblockN{Jingxuan Zhou\IEEEauthorrefmark{1},
Yinchao Yang\IEEEauthorrefmark{1},
Zhaohui Yang\IEEEauthorrefmark{2}\IEEEauthorrefmark{3},
Wei Xu\IEEEauthorrefmark{4},
Mohammad Shikh-Bahaei\IEEEauthorrefmark{1}}
 \IEEEauthorblockA{
   $\IEEEauthorrefmark{1}$Department of Engineering, King's College London, London, UK \\
   $\IEEEauthorrefmark{2}$College of Information Science and Electronic Engineering, Zhejiang University, Hangzhou, China\\
    $\IEEEauthorrefmark{3}$Zhejiang Provincial Key Laboratory of Info. Proc., Commun. \& Netw. (IPCAN), Hangzhou, China\\
     $\IEEEauthorrefmark{4}$School of Information, Science and Engineering, Southeast University, Nanjing, China\\
   E-mails: 
 jingxuan.zhou@kcl.ac.uk,
 yinchao.yang@kcl.ac.uk,
 yang\_zhaohui@zju.edu.cn,
 m.sbahaei@kcl.ac.uk,
 wxu@seu.edu.cn}}



\maketitle

\begin{abstract}
This paper proposes a fluid antenna (FA)-assisted near-field integrated sensing and communications (ISAC) system enabled by the extremely large-scale simultaneously transmitting and reflecting surface (XL-STARS). By optimizing the communication beamformer, the sensing signal covariance matrix, the XL-STARS phase shift, and the FA position vector, the Cram\'er-Rao bound (CRB), as a metric for sensing performance, is minimized while ensuring the standard communication performance. A double-loop iterative algorithm based on the penalty dual decomposition (PDD) and block coordinate descent (BCD) methods is proposed to solve the non-convex minimization problem by decomposing it into three subproblems and optimizing the coupling variables for each subproblem iteratively. Simulation results validate the superior performance of the proposed algorithm.
\end{abstract}


\section{Introduction}
The sixth generation (6G) mobile networks are poised to revolutionize emerging applications, including holographic video, the Metaverse, and digital twins \cite{xu2023edge, yang2023energy, zhang2019neural,olfat2008optimum, bobarshad2010low, nehra2010spectral, bobarshad2009m, shikh2007joint, nehra2010cross, ho2008design, shadmand2010multi, kobravi2007cross, fang2021secure, towhidlou2017improved, jia2020channel}. To satisfy the precipitous increase in key performance indicators (KPIs) of 6G networks, extensive research initiatives have been launched to propel innovation in cutting-edge wireless technologies. For instance, integrated sensing and communications (ISAC) has emerged as a promising solution that enhances spectrum and resource efficiency and facilitates the deployment of emerging applications, such as autonomous vehicles \cite{liu2022survey}. Furthermore, as a key enabler of future wireless communication and sensing systems, reconfigurable intelligent surfaces (RISs) introduce novel prospects for boosting both capacity and coverage of networks without additional power \cite{liu2021reconfigurable}. Additionally, millimeter-wave (mmWave) and terahertz (THz) multiple-input multiple-output (MIMO) technologies promote an exponential enhancement in peak data rates for 6G wireless communications by providing substantial spectral resources.


The development of the aforementioned emerging topics has significantly advanced the next-generation progress by catering to distinct scenarios with diverse KPIs. However, the realization of these scenarios typically necessitates the deployment of extremely large-scale antenna arrays (ELAAs) and the utilization of high-frequency bands. This leads to an increase of the Rayleigh distance $R = \frac{2D^2}{\lambda}$, where $D$ is the antenna aperture and $\lambda$ is the wavelength. As the Rayleigh distance determines the boundary between the near-field and the far-field region, receivers may locate in the near-field region instead of the far-field region of transmitters with ELAAs. Compared to far-field sensing based on planar wave models, near-field sensing utilizes spherical waves to estimate both the angles and distances of targets. This capability not only opens novel prospects for future wireless networks but also introduces new challenges. Therefore, intensifying research on near-field systems in existing 6G scenarios is imperative.

Moreover, although previous works highlight the pivotal role of RISs in ISAC systems, the simultaneously transmitting and reflecting surface (STARS) is proposed to compensate for the limitation of RISs that can only reflect signals. Compared to the conventional RIS, the STARS can provide $360^{\circ}$ full-space signal coverage \cite{liu2021star} by utilizing both sides. Therefore. the STARS is ideally suited for deployment in ISAC systems to partition space, allowing for the communications and sensing process to occur on both sides. Furthermore, new degrees of freedom (DoFs) offered by another emerging topic, fluid antennas (FAs), hold the potential to significantly enhance wireless communications and sensing performance \cite{zhu2023movable, ma2023mimo, chen2024joint}. Specifically, the real-time adjustment of FA positions allows adaptation to changing environmental conditions and varying task requirements. Also, in practice, the geometry of FA arrays can be pre-configured to facilitate flexible switching between communication and sensing functions in ISAC systems.



These hence motivate our work. Despite there have been some works about the STARS-aided wireless communications \cite{wang2023stars} and the FAs-enabled communications and sensing, the research field in FA-aided near-field sensing remains largely unexplored. Thus, in this paper, a FA-aided near-field ISAC system enabled by the XL-STARS is proposed. Our main contributions include: 1) an innovative near-field channel model based on the FA-aided near-field system (FANS) is proposed, 2) the Cram\'er-Rao bound (CRB) for mean square errors (MSE) of the target location $(r, \theta)$, i.e., the distance and angle of arrival (AoA), is derived, and an optimization problem is formulated to minimize the CRB by optimizing the communication beamformer, the sensing signal covariance matrix, the XL-STAR phase shift, and the FA position array while ensuring the desired communication performance, and 3) a dual-loop iterative algorithm based on penalty dual decomposition (PDD) and block coordinate descent (BCD) methods is proposed to solve the non-convex optimization problem by transforming the original problem into three solvable subproblems and iteratively optimize the coupling variables for each subproblem.

\section{System Model}
\begin{figure}[t!]
    \centering
    \includegraphics[width=75mm]{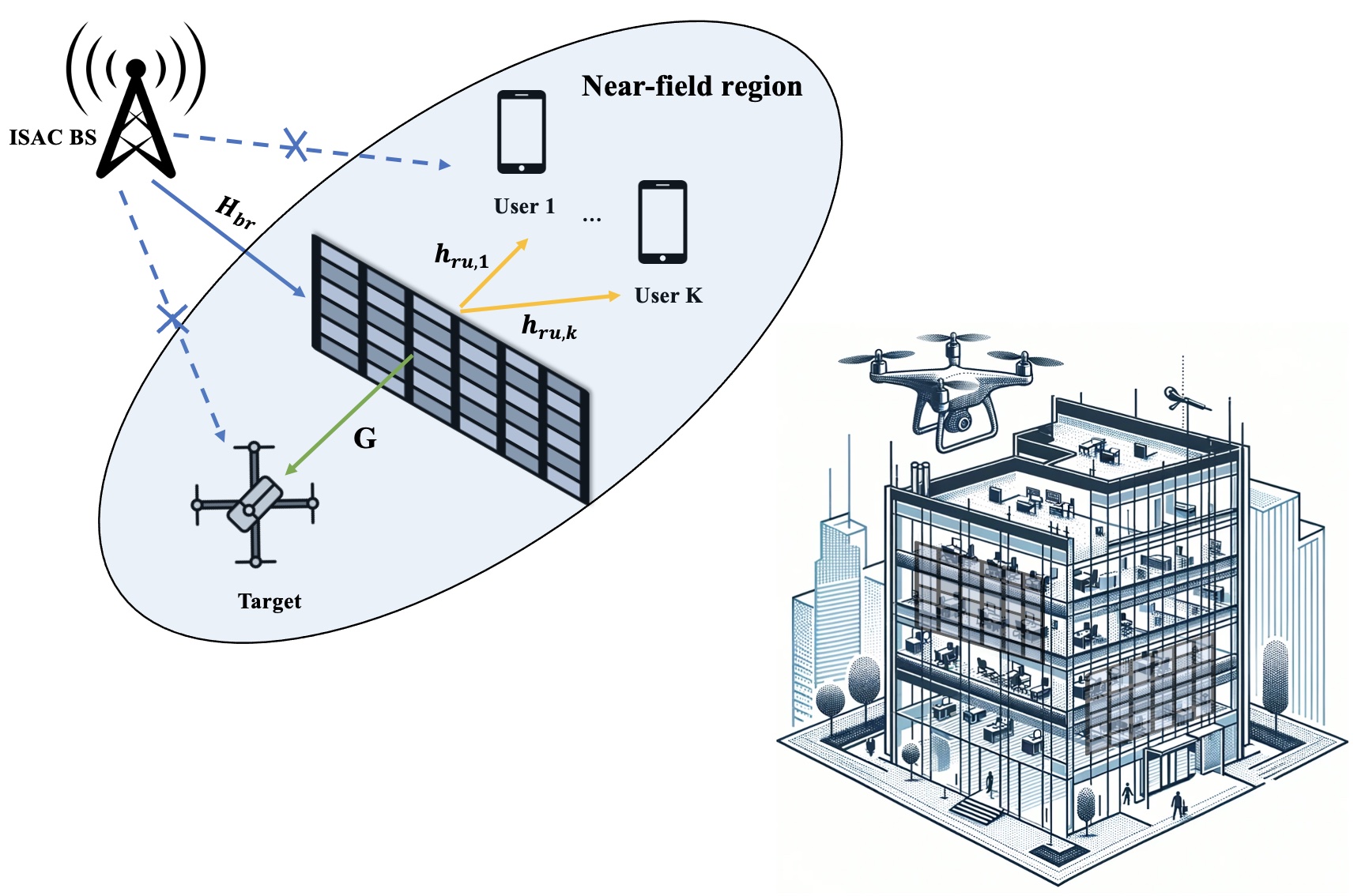}
    \caption{Near-field XL-STARS-enabled ISAC system}
    \label{scenario}
\end{figure}

As shown in Fig. \ref{scenario}, we envision a FA-aided near-field ISAC system operating in the mmWave frequency, where an XL-STARS with $N = 2\widetilde{N} + 1$ elements, denoted by set $\mathcal{N}$, locates in the far-field region of the base station (BS) with $M = 2\widetilde{M} + 1$ antennas. Further, we assume that a single-FA target and $K$ single-antenna users, referred to as set $\mathcal{K}$, are located in the near-field region of the XL-STARS. To fully explore the upper bound of the system performance, uniform linear arrays (ULAs) are considered to be equipped at both the BS and the XL-STARS. Also, we assume that all channel state information (CSI) is perfectly available to the BS and the XL-STARS in this paper.

\subsection{Signal Model}
We consider a coherent time block $T$, during which parameters for both the communication and sensing remain constant. Thus, the ISAC signal transmitted from the BS at time index $t\in[0,T]$ can be modeled as \begin{equation}
    \mathbf{x}[t] = \mathbf{x}_{0}[t] + \sum_{k=1}^{K}\mathbf{w}_{k}s_{k}[t],
    \vspace{-0.2cm}
\end{equation} 
where $\mathbf{w}_{k} \in \mathbb{C}^{M \times 1}$ denotes the fully digital transmit beamforming vector delivering information symbol $s_{k}[t]$ to user $k\in \mathcal{K}$. Communication signals are modeled as independent Gaussian random variables with zero mean and standard power. Further, dedicated sensing signal $\mathbf{x}_{0}[t] \in \mathbb{C}^{M \times 1}$ is synthesized using pseudo-random coding to ensure that $\mathbb{E}\left [ \mathbf{s}[t] \mathbf{s}^{H}[t] \right ] = \textbf{I}_K$ and $\mathbb{E}\left [ \mathbf{s}[t] \mathbf{x_0}^{H}[t] \right ] = \textbf{0}_{K\times M}$. Also, as signal $\mathbf{x}_{0}[t]$ employs multiple beam transmissions, its covariance matrix $\mathbf{R}_{0} = \mathbb{E}\left [ \mathbf{x}_{0}[t] \mathbf{x}_{0}^{H}[t] \right ]$ is of general rank. Thus, the covariance matrix for the transmitted signal can be given by $\mathbf{R}_{x} = \mathbb{E}\left [ \mathbf{x}[t] \mathbf{x}^{H}[t] \right ] = \sum_{k=1}^{K}\mathbf{w}_{k} \mathbf{w}_{k}^{H} + \mathbf{R}_{0}$.

\subsubsection{Communication Model}

We define matrices of the XL-STARS transmission coefficients (TCs) and reflection coefficients (RCs) as $\mathbf{\Phi}_{t} \in \mathbb{C}^{N \times N}$ and $\mathbf{\Phi}_{r} \in \mathbb{C}^{N \times N}$, respectively. The channel from the BS to the XL-STARS and from the XL-STARS to user $k$ are denoted as $\mathbf{H}_{br} \in \mathbb{C}^{N \times M}$ and $\mathbf{h}_{ru,k} \in \mathbb{C}^{N \times 1}$, respectively. As links between the BS to users/target are considered to be obstructed, the effective channel from the BS to user $k$ is given by cascaded channel $\mathbf{h}_{u,k} \triangleq \mathbf{h}_{ru,k}^{H} \mathbf{\Phi}_{t} \mathbf{H}_{br}$. Therefore, by denoting the additive white Gaussian noise (AWGN) for user $k$ as $n_{k}[t] \sim \mathcal{CN}(0,\sigma _{k}^{2})$, the signal received by user $k$ at time $t$ can be modeled as \begin{equation} \label{eqq1}
\footnotesize
    y_{k}[t] = \mathbf{h}_{u,k}\mathbf{w}_{k}s_{k}[t] + \sum_{i\neq k} \mathbf{h}_{u,k} \mathbf{w}_{i}s_{i}[t] + \mathbf{h}_{u,k} \mathbf{x}_{0}[t]+ n_{k}[t].
    \vspace{-0.2cm}
\end{equation}

Also, we consider the signal-to-interference-plus-noise ratio (SINR) as the performance metric for communications. According to \eqref{eqq1}, the SINR for user $k$ can be expressed as \begin{equation}
\gamma_{k}= \frac{\left |\mathbf{h}_{u,k}\mathbf{w}_{k} \right |^{2}}{\sum_{i \neq k}\left | \mathbf{h}_{u,k} \mathbf{w}_{i}  \right |^{2} + \mathbf{h}_{u,k} \mathbf{R}_{0}\mathbf{h}_{u,k}^H + \sigma_{k}^{2}}.
\end{equation}

\subsubsection{Sensing Model}


As the number of the target is fewer than the number of the XL-STARS elements, the space-division multiple access (SDMA) can be used to achieve new DoFs. We assume that the FA is connected to the radio frequency (RF) chain via a flexible cable; hence, the FA can be moved within a local region, which is considered as a line with $Q = 2\widetilde{Q} + 1$ movable positions (MPs) in this paper. Positions activation of the FA is denoted by $u_{q}, \forall q \in \{-\widetilde{Q},...,\widetilde{Q}\}$. The active position is denoted by $u_{q} = 1$, while $u_{q} = 0$ indicates the inactive positions. Then, received echo signal $\mathbf{Y}_{s} \in \mathbb{C}^{N \times (T+1)}$ at the BS over time block $T$ can be modeled as \begin{equation} \label{eq4}
    \mathbf{Y}_{s} = \sum_{q=-Q}^{Q}u_{q} \mathbf{G}_q \mathbf{\Phi}_{r} \mathbf{H}_{br} \mathbf{X} + \mathbf{N}_{s}, u_{q} \in \{0,1\},\vspace{-0.1cm}
\end{equation} where $\mathbf{G}_q \in \mathbb{C}^{N \times N}$ is the near-field round-trip channel matrix, $\mathbf{X} \in \mathbb{C}^{M \times (T+1)}$ denotes the transmitted signal over time block $T$ and $\mathbf{N}_{s}$ represents the AWGN noise with each element following $\mathcal{CN}(0,\sigma _{s}^{2})$. Notably, the echo signal is captured within a designated range-Doppler bin, hence excluding range and Doppler details.

\subsection{STAR Model}

Let $\alpha_{i,n} \in [0,1]$ and $\varphi_{i,n} \in [0,2\pi]$ represent the amplitude and phase-shift response of the $n$-th XL-STARS element with values determined by the performance of element $n$. Hence, TCs $\mathbf{\Phi}_{t}$ and RCs $\mathbf{\Phi}_{r}$ of the XL-STARS can be modeled as $\mathbf{\Phi}_{i} = \Diag (\alpha_{i,1}e^{j\varphi _{i,1}},...,\alpha_{i,N}e^{j\varphi _{i,N}}), \forall i \in \{t,r\}$. Assuming that phase shifts of the TC and the RC can be independently adjusted, the amplitudes need to satisfy that \begin{equation}
    \alpha_{t,n}^{2} + \alpha_{r,n}^{2} = 1, \forall n \in \mathcal{N}.
\end{equation}

\subsection{Channel Model}

We consider a two-dimensional Cartesian coordinate system, where the ULA center of the XL-STARS is the origin and user $k$ is allocated at location $(r_k, \theta_k)$. Moreover, we denote the distance and AoA from the FA position $q$ to the origin as $r_{q}$ and $\theta_{q}$, respectively. Thus, the distance from element $n$, $\forall n\in\{-\widetilde{N},...,\widetilde{N}\}$, of the XL-STARS to user $k$/the target can be calculated as \begin{equation}
    r_{n,i}\left ( r_{i},\theta_{i}  \right ) = \sqrt{r_{i}^{2} + n^{2}d^{2} - 2r_{i}nd\cos\theta_{i}}, \forall i \in \{k,q\},
\end{equation} where $d$ represents the element spacing for the XL-STARS. Further, based on properties of trigonometric functions, $r_{q}$ and $\theta_q$ can be obtained by $r_{q} = \sqrt{r_{0}^{2} + q^{2}d_1^{2} - 2r_{0}qd_1\cos\theta_{0}}$ and $ \theta_{q} = \arcsin (\frac{r_{0}\sin\theta_{0}}{r_{q}})$, respectively, where $r_0$, $\theta_0$, and $d_1$ represent the distance and AoA from the MPs array center of the FA to the origin, and the MPs spacing, respectively. Then, by denoting the near-field beamforming vector as $\mathbf{a}(r_{i}, \theta_{i})$, the $n$-th element of $\mathbf{a}(r_{i}, \theta_{i})$ can be given by \begin{equation}
    [\mathbf{a}\left ( r_{i},\theta_{i} \right )]_{n} = e^{-j\frac{2\pi}{\lambda}(r_{n,i}(r_{i},\theta_{i}) - r_{i})}, \forall i \in \{k,q\}.
    \vspace{-0.1cm}
\end{equation}

Next, let $\beta_{i}$ denote the complex channel gain, and thus the channel from element $n$ to user $k$/the target can be obtained by $ \mathbf{h}_{n,i}\left ( r_{i},\theta_{i} \right ) = \beta_{i}[\mathbf{a}\left ( r_{i},\theta_{i} \right )]_n, \forall i \in \{k,q\}$. Thus, channel vector $\mathbf{h}_{ru,k}$ between the XL-STARS and user $k$ is given by \begin{equation}
    \mathbf{h}_{ru,k} = \beta_{k} \mathbf{a}\left (r_{k},\theta_{k} \right ),
\end{equation} and near-field channel matrix $\mathbf{G}_q$ can be modeled as \begin{equation} \label{g}
    \mathbf{G}_q = \beta_{q} \mathbf{a}(r_{q},\theta_{q}) \mathbf{a}^{T}(r_{q},\theta_{q}).
\end{equation}


Finally, far-field channel $\mathbf{H}_{br}$ between the BS and the XL-STARS is modeled as \begin{equation}
   \mathbf{H}_{br} = \beta_{s} \mathbf{b}(\theta_{s})\mathbf{b}^{T}(\theta_{s}),
\end{equation} where $[\mathbf{b}(\theta_s)]_{n} = e^{-j\frac{2\pi}{\lambda}(-md_{s}\cos\theta_{s})}, \forall m\in\{-\widetilde{M},...,\widetilde{M}\}$ denotes the $n$-th element of far-field beamforming vector $\mathbf{b}(\theta_s)$. Additionally, $\beta_s$, $\theta_{s}$, and $d_{s}$ denote the path loss of channel $\mathbf{H}_{br}$, the AoA of the XL-STARS relative to the BS, and the antenna spacing of the BS, respectively.

\subsection{Sensing Performance Matrices - CRB}

The near-field sensing aims to estimate the target location from echo signals over time block $T$. Generally, sensing performance is assessed by comparing MSE between estimated and actual values. However, obtaining MSE in closed form is generally hard and the minimization problem of MSE is intractable, and thus MSE cannot be used as the metric of sensing performance in this case. As a remedy, the CRB can provide a theoretical lower bound of MSE when an unbiased estimator is used \cite{kay1993fundamentals}. The CRB derivation begins with the vectorization of \eqref{eq4}, i.e., $\mathbf{y}_{s} = \text{vec}(\mathbf{Y}_{s}) = \mathbf{v} + \mathbf{n}_{s}$, where $\mathbf{v} = \text{vec}(\mathbf{G}_q\mathbf{\Phi}_{r} \mathbf{H}_{br} \mathbf{X})$ and $\mathbf{n}_{s} = \text{vec}(\mathbf{N}_{s})$. Then, let $\mathbf{\xi} = \left [ r_{q}, \theta_{q}, \beta_q^{r}, \beta_q^{i} \right ]$ denote the unknown parameters, where $\beta_q^{r} = \Re\{\beta_q\}$ and $\beta_q^{i} = \Im\{\beta_q\}$. The Fisher information matrix (FIM) for estimating vector $\mathbf{\xi}$ is \begin{equation}
\mathbf{J_{\xi }} = \begin{bmatrix}
\mathbf{J}_{11,q} & \mathbf{J}_{12,q}\\
\mathbf{J}_{12,q}^{T} & \mathbf{J}_{22,q}
\end{bmatrix}.
\end{equation} Next, the CRB matrix can be given by\vspace{-0.2cm} \begin{equation}
    CRB(r_{q},\theta_{q}) = \sum_{q=-Q}^{Q}u_{q}[\mathbf{J}_{11,q} - \mathbf{J}_{12,q} \mathbf{J}_{22,q}^{-1} \mathbf{J}_{12,q}^{T}]^{-1},
\end{equation} where $\mathbf{J}_{11,q} = \begin{bmatrix}
J_{r_{q}r_{q}} & J_{r_{q}\theta_{q}} \\ 
J_{r_{q}\theta_{q}}& J_{\theta_{q}\theta_{q}} 
\end{bmatrix}$, $\mathbf{J}_{12,q} = \begin{bmatrix}
J_{r_{q}\beta_q^{r}} & J_{r_{q}\beta_q^{i}} \\ 
J_{\theta_{q}\beta_q^{r}}& J_{\theta_{q}\beta_q^{i}}\end{bmatrix}$, and $\mathbf{J}_{22,q} = \begin{bmatrix}
J_{\beta_q^{r}\beta_q^{r}} & 0 \\ 
0 & J_{\beta_q^{i}\beta_q^{i}}
\end{bmatrix}$. The value of each entry is derived in Appendix \ref{ap1}.

\addtolength{\topmargin}{0.03in}

\section{Problem Formulation and Proposed Solution} 
\subsection{Problem Formulation}

As diagonal entries of the CRB matrix represent minimum variances for estimated parameters, we aim to minimize the trace of the matrix, i.e., $\Tr(\mathbf{J}^{-1})$, where $\mathbf{J}^{-1}$ is the inverse of the FIM and $ \mathbf{J}^{-1} = [\mathbf{J}_{11,q} - \mathbf{J}_{12,q} \mathbf{J}_{22,q}^{-1} \mathbf{J}_{12,q}^{T}]^{-1} \succeq 0$. According to \cite{boyd2004convex}, $\Tr(\mathbf{A}^{-1})$ is a matrix decreasing on the positive semidefinite matrix space. Hence, based on the Schur complement condition \cite{zhang2006schur}, by introducing an auxiliary matrix $\mathbf{U} \in \mathbb{C}^{2 \times 2}$, which satisfies $\mathbf{J} \succeq \mathbf{U} \succeq 0$, the CRB$(r_{q},\theta_{q})$ minimization problem can be formulated as \begin{subequations} \label{e2}
\begin{align}
\min_{\mathbf{U},\mathbf{w}, \mathbf{R}_{0}, \mathbf{o}_r, \mathbf{o}_t, \mathbf{u}}\; & \Tr(\mathbf{U}^{-1})\\
\textrm{s.t.} \quad & \sum_{q=-Q}^{Q}u_{q}\begin{bmatrix}
\mathbf{J}_{11,q} - \mathbf{U} & \mathbf{J}_{12,q}\\
\mathbf{J}_{12,q}^{T} & \mathbf{J}_{22,q}
\end{bmatrix}\succeq 0, \label{23b} \\ 
& \sum_{q=-Q}^{Q}u_{q}=1, \forall q,\label{23c} \\
& u_{q} \in \{0,1\}, \forall q, \label{23d}\\
& \gamma_{k} \geq \overline{\gamma}_{k}, \forall k, \label{23e}\\
& \Tr(\mathbf{w}\mathbf{w}^{H} + \mathbf{R}_{0}) \leq P,\label{23f}\\
& \alpha_{t,n}^{2} + \alpha_{r,n}^{2} = 1, 0 \leq \alpha_{t,n}, \alpha_{r,n} \leq 1, \forall n, \label{23g}\\
& \mathbf{R}_{0} \succeq 0, \mathbf{U} \succeq 0, \label{23h}  
\end{align}
\end{subequations}
where $\overline{\gamma}_{k} \geq 0$ is the SINR threshold for user $k$, $P \geq 0$ denotes the power budget, $\mathbf{w} = \left [ \mathbf{w}_{1}, ..., \mathbf{w}_{k} \right ]$ represents the transmit beamforming matrix, $\mathbf{o}_{i} =[\alpha_{i,1}e^{j\varphi_{i,1}},...,\alpha_{i,N}e^{j\varphi_{i,N}}], \forall i \in \{t,r\}$ is the vector formed from diagonal elements of $\mathbf{\Phi}_{i}$, and $\mathbf{u} = \left [u_{1},...,u_{q}\right ]$ is the antenna position vector (APV). Further, the threshold of communication performance and the power budget are constrained in \eqref{23e} and \eqref{23f}, respectively. The amplitudes of the $n$-th XL-STARS element need to satisfy constraint \eqref{23g}. Finally, constraints \eqref{23b}, \eqref{23c}, and \eqref{23d} guarantee the performance of the FA. 


\subsection{PDD-Based Iterative Algorithm}

Problem \eqref{e2} is non-convex due to the discrete and non-convex constraints. Thus, an iterative double-loop algorithm based on the PDD and BCD methods is proposed \cite{shi2020penalty}, where problem \eqref{e2} is transferred to three augmented Lagrangian (AL) problems and solved in the inner loop of the algorithm, while Lagrangian dual variable and penalty terms are updated in the outer loop. Hence, to solve problem \eqref{e2} by using the aforementioned algorithm, we first construct an AL problem for problem \eqref{e2} that allows each BCD step to have a simple solution. By introducing an auxiliary variable $\Lambda$, problem \eqref{e2} can be transformed to \begin{subequations} \label{e3}
\begin{align}
\min_{\mathbf{V}}\quad & \Tr(\mathbf{U}^{-1})\\
\textrm{s.t.} \quad & \Lambda  = \mathbf{\Phi}_{r} \mathbf{H}_{br} \mathbf{R}_{x} \mathbf{H}_{br}^{H} \mathbf{\Phi}_{r}^{H}, \label{25b}\\
&  \sum_{q=-Q}^{Q}u_{q}\begin{bmatrix}
\mathbf{J}_{11,q}(\Lambda ) - \mathbf{U} & \mathbf{J}_{12,q}(\Lambda )\\
\mathbf{J}_{12,q}(\Lambda )^{T} & \mathbf{J}_{22,q}(\Lambda )
\end{bmatrix}\succeq 0, \label{25c} \\ 
& \eqref{23c} - \eqref{23h}, \notag
\end{align}
\end{subequations} 
where $\mathbf{V} = \{\mathbf{U},\mathbf{w}, \mathbf{R}_{0}, \Lambda, \mathbf{o}_{t}, \mathbf{o}_{r}, \mathbf{u}\}$. Next, by incorporating Lagrangian dual variable $\Gamma$ and penalty factor $\varrho \geq 0$, the AL problem for \eqref{e2} is constructed as\begin{subequations} \label{e4}
\begin{align}
\min_{\mathbf{V}}\quad & \Tr(\mathbf{U}^{-1}) + \frac{1}{2\varrho}\left \| \Lambda  - \mathbf{\Phi}_{r} \mathbf{H}_{br} \mathbf{R}_{x} \mathbf{H}_{br}^{H} \mathbf{\Phi}_{r}^{H} + \varrho\Gamma\right \|^{2} \label{e4a}\\ 
\textrm{s.t.} \quad & \eqref{23c} - \eqref{23h}, \eqref{25c}. \notag
\end{align}
\end{subequations}

\vspace{-0.4cm}
\begin{algorithm} \label{a1}
    \caption{Penalty Dual Decomposition-Based Algorithm}
    \begin{algorithmic}[1]
        \STATE Initialize $\mathbf{V}^{0}$, $\Gamma^{0}$, $\varrho^{0} > 0$, $0<z<1$, $t = 1$, and set two sequences $\{\eta^{t} > 0\}$ and $\{\epsilon^{t} > 0\}$.
        \REPEAT
        \STATE Solve \eqref{e4} to obtain $\mathbf{V}^{t}$.
        \IF{$\left \| h (\mathbf{V}^{t})\right \|_{\infty } \leq \eta^{t}$}
        \STATE{$\Gamma^{t+1} = \Gamma^{t} + \frac{1}{\varrho}(\Lambda^{t+1}  - \mathbf{\Phi}_{r}^{t+1} \mathbf{H}_{br} \mathbf{R}_{x}^{t+1} \mathbf{H}_{br}^{H}(\mathbf{\Phi}_{r}^{t+1})^{H})$};\\{$\varrho^{t+1} = \varrho^{t}$}.
        \ELSE
        \STATE{$\Gamma^{t+1} = \Gamma^{t}$; $\varrho^{t+1} = z\varrho^{t}$}.
        \ENDIF
        \STATE {$t = t + 1$}.
        \UNTIL{$h (\mathbf{V}^{t})$ falls below a predefined threshold.}
    \end{algorithmic}
\end{algorithm}

\textbf{Algorithm 1} demonstrates the outer loop of the proposed algorithm based on the PDD method, where $h (\mathbf{V}^{t})$ represents the constraint violation function and $\eta^{t} = 0.99h(\mathbf{V}^{t})$ is a sequence converging to zero. 
When penalty factor $\varrho$ is sufficiently small, observations can be made that the value of the constraint violation can approach zero. This satisfies equality constraint \eqref{25b} and penalty term $\frac{1}{2\varrho}\left \| \Lambda  - \mathbf{\Phi}_{r} \mathbf{H}_{br} \mathbf{R}_{x} \mathbf{H}_{br}^{H} \mathbf{\Phi}_{r}^{H} + \varrho\Gamma\right \|^{2}$. For simplicity, we let $U(\mathbf{V})$ refer to the objective function $\Tr(\mathbf{U}^{-1}) + \frac{1}{2\varrho}\left \| \Lambda  - \mathbf{\Phi}_{r} \mathbf{H}_{br} \mathbf{R}_{x} \mathbf{H}_{br}^{H} \mathbf{\Phi}_{r}^{H} + \varrho\Gamma\right \|^{2}$ in the following text.

\subsection{Proposed BCD Algorithm}

In the inner loop, optimization variable set $\mathbf{V}$ is divided into three blocks, i.e., $\{\mathbf{U}, \Lambda , \mathbf{w}, \mathbf{R}_{0}\}$, $\{\mathbf{o}_{t}, \mathbf{o}_{r}\}$, and $\{\mathbf{u}\}$. Next, we use the BCD method to optimize variables for each block by keeping other blocks constant. Thus, problem \eqref{e4} is decomposed into the following three subproblems.

\subsubsection{Subproblem With Respect to $\{\mathbf{U}, \Lambda, \mathbf{w}, \mathbf{R}_{0}\}$}

For any given $\{\mathbf{u}\}$ and $\{\mathbf{o}_{t}, \mathbf{o}_{r}\}$, the subproblem is given by \begin{subequations} \label{e5}
\begin{align}
\min_{\mathbf{U}, \Lambda, \mathbf{w}, \mathbf{R}_{0}}\quad & U(\mathbf{V})\\
\textrm{s.t.} \quad & \begin{bmatrix}
\mathbf{J}_{11,q}(\Lambda) - \mathbf{U} & \mathbf{J}_{12,q}(\Lambda)\\
\mathbf{J}_{12,q}(\Lambda)^{T} & \mathbf{J}_{22,q}(\Lambda)
\end{bmatrix}\succeq 0, \label{27b} \\ 
& \eqref{23e}, \eqref{23f}, \eqref{23h}.\notag
\end{align}
\end{subequations}

To solve problem \eqref{e5}, we first convert it into a semidefinite programming (SDP) problem and then solve it using the semidefinite relaxation (SDR) approach. It can be easily proved that given $\mathbf{w} =\left [ \mathbf{w}_{1},...,\mathbf{w}_{K}\right ]$, there exists $\mathbf{R}_{x}$ and $\mathbf{R}_{0}$ satisfying $\mathbf{R}_{x}=\sum_{k=1}^{K}\mathbf{w}_{k}\mathbf{w}_{k}^{H} + \mathbf{R}_{0}$ if and only if $\mathbf{R}_{x} \succeq \sum_{k=1}^{K}\mathbf{w}_{k} \mathbf{w}_{k}^{H} = \mathbf{w}\mathbf{w}^{H}$. Thus, by defining an auxiliary variable $\mathbf{\Omega}_{k} = \mathbf{w}_{k} \mathbf{w}_{k}^{H}, \forall k \in \mathcal{K}$ that satisfy $\mathbf{\Omega}_{k}\succeq 0$ and $\text{rank}(\mathbf{\Omega}_{k}) = 1$, the SDP problem of \eqref{e5} can be given by
\begin{subequations} \label{e6}
\begin{align}
\min_{\mathbf{U}, \Lambda, \mathbf{R}_{0}, \mathbf{\Omega}_{k}}\quad & U(\mathbf{V}) \\
\textrm{s.t.} \quad & (1+\frac{1}{\overline{\gamma}_{k}}) \mathbf{h}_{u,k}^{H}\mathbf{\Omega}_{k} \mathbf{h}_{u,k} \geq \mathbf{h}_{u,k}^{H} \mathbf{R}_{x} \mathbf{h}_{u,k} + \sigma _{k}^{2}, \forall k, \label{28b}\\
& \mathbf{R}_{x} \succeq \sum_{k\in \mathcal{K}} \mathbf{\Omega}_{k}, \mathbf{\Omega}_{k} \succeq 0, \forall k, \label{28d} \\
& \Tr(\mathbf{R}_{x}) \leq P, \\
& \eqref{23h}, \eqref{27b}.\notag
\end{align}
\end{subequations}

Although problem \eqref{e6} can be easily solved by the existing solver 
after relaxing the rank-one constraint $\text{rank}(\mathbf{\Omega}_{k}) = 1$, the solution might with a higher rank. However, according to \cite{luo2010semidefinite}, given an arbitrary high-rank global optimal solution (GOS), a rank-one GOS for the original problem can always be reconstructed by 
\begin{equation}
    \mathbf{R}_{x}^{*} = \widehat{\mathbf{R}}_{x}, \mathbf{w} _{k}^{*} = (\mathbf{h}_{u,k}^{H}\widehat{\mathbf{\Omega}} _{k} \mathbf{h}_{u,k})^{-\frac{1}{2}}\widehat{\mathbf{\Omega}}_{k} \mathbf{h}_{u,k},
\end{equation} and then the GOS for the dedicated sensing signal covariance matrix $\mathbf{R}_{0}$ can be obtained by $\mathbf{R}_{0} = \mathbf{R}_{x}^{*} - \sum_{k \in \mathcal{K}}\mathbf{w}_{k}^{*}(\mathbf{w}_{k}^{*})^{H}$.


\subsubsection{Subproblem With Respect to $\{\mathbf{o}_{t}, \mathbf{o}_{r}\}$}

By dropping the constant part of problem \eqref{e4}, for any given $\{\mathbf{u}\}$ and fixed $\{\mathbf{U},\Lambda, \mathbf{w}, \mathbf{R}_{0}\}$, the subproblem is given by
\begin{subequations} \label{e7}
\begin{align}
\min_{\mathbf{o}_{t}, \mathbf{o}_{r}}\quad & \left \| \Lambda - \mathbf{\Phi}_{r} \mathbf{H}_{br} \mathbf{R}_{x} \mathbf{H}_{br}^{H} \mathbf{\Phi}_{r}^{H} + \varrho\Gamma\right \|^{2}\\
\textrm{s.t.} \quad 
& \eqref{23e}, \eqref{23g}. \notag
\end{align}
\end{subequations}

For solving problem \eqref{e7}, we first define $\mathbf{\mathcal{R}}_x = \mathbf{H}_{br} \mathbf{R}_{x} \mathbf{H}_{br}^{H}$. The eigenvalue decomposition of matrix $\mathbf{\mathcal{R}}_x$ is given by $\mathbf{\mathcal{R}}_x = \sum_{j=1}^{R} \rho_j \mathbf{v}_{j}\mathbf{v}_{j}^{H} = \mathbf{\widehat{v}}_{j}\mathbf{\widehat{v}}_{j}^{H}$, where $\mathbf{\widehat{v}}_{j} = \sqrt{\rho_j}\mathbf{v}_{j}$ with $\rho_j$ and $\mathbf{v}_{j}$ as the eigenvalue and the corresponding eigenvector, respectively. Next, by introducing auxiliary variables $\mathbf{\Theta}_{k,i} = \mathbf{\mathcal{H}}_k\mathbf{w} _{i}\mathbf{w} _{i}^{H}\mathbf{\mathcal{H}}_k^{H}$ and $\mathbf{\Upsilon}_k  = \mathbf{\mathcal{H}}_k \mathbf{R}_{s}\mathbf{\mathcal{H}}_k^{H}$, where $\mathbf{\mathcal{H}}_k = \Diag(\mathbf{h}_{ru,k}^{H}) \mathbf{H}_{br}$, problem \eqref{e7} can be transformed to \begin{subequations} \label{e8}
\begin{align}
\min_{\mathbf{o}_{t}, \mathbf{o}_{r}}\quad & \left \| \widehat{\Lambda} - \sum_{j=1}^{R}\mathbf{\widehat{V}}_{j}\mathbf{o}_{r}\mathbf{o}_{r}^{H}\mathbf{\widehat{V}}_{j}^{H} + \varrho\Gamma\right \|^{2}\\
\textrm{s.t.} \quad & \frac{1}{\overline{\gamma }_{k}}\mathbf{o}_{t}^{H}\mathbf{\Theta} _{k,k}^{*}\mathbf{o}_{t}\geq \sum _{i\neq k}\mathbf{o}_{t}^{H}\mathbf{\Theta}  _{k,i}^{*}\mathbf{o}_{t} + \mathbf{o}_{t}^{H}\mathbf{\Upsilon}_k^{*}\mathbf{o}_{t} + \sigma_{k}^{2}, \forall k, \label{37b}\\
& \left | [\mathbf{o}_{t}]_{n} \right | ^{2}+ \left | [\mathbf{o}_{r}]_{n} \right | ^{2} = 1, \forall n, \label{37c}
\end{align}
\end{subequations}
where $\mathbf{\widehat{V}}_{j}  = \Diag(\mathbf{\widehat{v}}_{j})$ and $\widehat{\Lambda} = \Lambda + \varrho\Gamma$. Here, problem \eqref{e8} is in the quadratic form relative to optimizing variables $\mathbf{o}_{t}$ and $\mathbf{o}_{r}$, and thus can be solved by the SDR approach. By defining variable $\widehat{\mathbf{\Phi}}_{i} = \mathbf{o}_{i}\mathbf{o}_{i}^{H}, \forall i \in \{t,r\}$, the SDP problem of \eqref{e8} can be formulated as \begin{subequations} \label{e9}
\begin{align}
\min_{\widehat{\mathbf{\Phi}}_{t},\widehat{\mathbf{\Phi}}_{r}}\quad & \left \| \widehat{\Lambda} - \sum_{j=1}^{R}\mathbf{\widehat{V}}_{j}\widehat{\mathbf{\Phi}}_{r}\mathbf{\widehat{V}}_{j}^{H} + \varrho\Gamma\right \|^{2}\\
\textrm{s.t.} \quad & \frac{1}{\overline{\gamma}_{k}}\Tr(\mathbf{\Theta} _{k,k}^{*}\widehat{\mathbf{\Phi}}_{t})\geq \nonumber \\
&\sum _{i\neq k}\Tr(\mathbf{\Theta}_{k,i}^{*}\widehat{\mathbf{\Phi}}_{t})+ \Tr(\Upsilon_{k}^{*} \widehat{\mathbf{\Phi}}_{t})+ \sigma _{k}^{2}, \forall k, \label{38b}\\
& \left | [\widehat{\mathbf{\Phi}}_{t}]_{n} \right | ^{2}+ \left | [\widehat{\mathbf{\Phi}}_{r}]_{n} \right | ^{2} = 1, \forall n, \label{38c}\\
& \widehat{\mathbf{\Phi}}_{t} \succeq 0, \widehat{\mathbf{\Phi}}_{r} \succeq 0, \label{38d}
\end{align}
\end{subequations} where the rank-one constraint is relaxed. The solution of problem \eqref{e9} can be easily obtained with an existing solver but might be with a higher rank. However, similar to problem \eqref{e6}, the rank-one GOS of the original problem \eqref{e7} can always be reconstructed from any higher-rank GOS of \eqref{e9} via the eigenvalue decomposition or the Gaussian randomization approach. 

\subsubsection{Subproblem With Respect to $\{\mathbf{u}\}$ }For the fixed $\{\mathbf{o}_{t}, \mathbf{o}_{r}\}$ and $\{\mathbf{U}, \Lambda, \mathbf{w}, \mathbf{R}_{0}\}$, the subproblem is given by
\begin{subequations} \label{e10}
\begin{align}
\min_{\mathbf{u}}\quad & U(\mathbf{V}) \\
\textrm{s.t.} \quad &\eqref{23b}, \eqref{23c}. \notag
\end{align}
\end{subequations}

To solve problem \eqref{e10}, discrete variables in \eqref{23c} need to be relaxed to continuous variables, thereby transforming problem \eqref{e10} into\begin{subequations} \label{e11}
\begin{align}
\min_{\mathbf{u}}\quad & U(\mathbf{V}) \\
\textrm{s.t.} \quad & 0 \leq u_{q} \leq 1, \forall q, \\
& \eqref{23b}, \eqref{23c}. \notag
\end{align}
\end{subequations} 

Problem \eqref{e11} is a standard linear programming problem that can be easily solved, where the integer solution of the original problem \eqref{e10} can be obtained by using the rounding method.
\begin{algorithm} 
    \caption{Block Coordinate Descent-Based Algorithm} \label{a2}
    \begin{algorithmic}[1]
        \STATE Set the initial $\mathbf{V}^{(0)} = (\mathbf{C}^{(0)}, \mathbf{O}^{(0)}, \mathbf{u}^{(0)})$, the iteration number $t = 0$, and the maximal iteration number $T_{max}$;
        \STATE Compute the objective value $V^{(0)}_{obj} = U(\mathbf{V}^{(0)})$;\REPEAT
        \STATE Set $t = t + 1$.
        \STATE Obtain optimal $\mathbf{C}^{(t)}$ of \eqref{e6} with fixed $(\mathbf{O}^{(t-1)}, \mathbf{u}^{(t-1)})$;\STATE Obtain optimal $\mathbf{O}^{(t)}$ of \eqref{e9} with fixed $(\mathbf{C}^{(t)}, \mathbf{u}^{(t-1)})$;
        \STATE Obtain optimal $\mathbf{u}^{(t)}$ of \eqref{e11} with fixed $(\mathbf{C}^{(t)}, \mathbf{O}^{(t)})$;
        \UNTIL{the fractional reduction of the objective value falls below a predefined threshol or $t > T_{max}$.}
    \end{algorithmic}
\end{algorithm}
Moreover, the inner algorithm based on the BCD method for solving problem \eqref{e4} is shown in \textbf{Algorithm} \ref{a2}, where $\mathbf{C} = \{\mathbf{U}, \Lambda, \mathbf{w}, \mathbf{R}_{0}\}$ and $\mathbf{O} = \{\mathbf{o}_{t}, \mathbf{o}_{r}\}$. Generally, \textbf{Algorithm} \ref{a2} ensures convergence to a stationary point 
within the polynomial time when the algorithm is employed to update block $\{\mathbf{u}\}$ \cite{razaviyayn2013unified}. The proof for the convergence of \textbf{Algorithm} \ref{a2} is shown in Appendix \ref{ap2}.
Furthermore, the main computational complexity of \textbf{Algorithm} \ref{a2} stems from solving QSDP problems \eqref{e6} and \eqref{e8}. Given a solution accuracy $\epsilon$, the worst case complexity to solve QSDP problems \eqref{e6} and \eqref{e8} with the primal dual interior-point method are $O((K^{6.5}M^{6.5} + N^{6.5})\log(1/\epsilon))$ and $O((K+N)^{6.5}N^{6.5}\log(1/\epsilon))$, respectively \cite{toh2008inexact}. 
\vspace{-0.4cm}

\section{Numerical Results}
\vspace{-0.4cm}
In this section, simulation results are presented to verify the effectiveness of the proposed FANS aided by the XL-STARS. The ULAs of BS and XL-STARS are assumed to have $M = 11$ and $N = 201$ half-wavelength spaced antenna/elements, respectively. With the carrier frequency $f_c = 10$ GHz, the Rayleigh distance of the XL-STARS is extended to $R = 600$ m. As we consider that users/the target are in the near-field region of the XL-STARS, users are randomly scattered within a range of $(20$ m$-40$ m$, 30^{\circ}-120^{\circ})$ at the transmission side of the XL-STARS, while the target locates in location $(20$ m$, 30^{\circ})$ at the reflective side of the XL-STARS \cite{wang2023stars}. The BS is considered to be located in $180$ m and $30^{\circ}$ from the XL-STARS. Further, the FA of the target has $Q = 3$ MPs with a third-wavelength MPs spacing. All channels are considered to be based on the Rician channel model with the path loss given by $\beta_i = \beta_{0}+20\log_{10}(d)$ dB, $\forall i \in \{k, q, s\}$, where $\beta_{0} = 30$ dB at $d = 1$ m. Moreover, we set the noise power at users/the target as $-110$ dB and the coherent time block as $T = 100$. Simulation results are compared with a near-field ISAC system \cite{wang2023near} (Benchmark 1), a far-field STAR-aided ISAC system \cite{wang2023stars} (Benchmark 2), and the proposed system enabled by fixed-position antenna (FPA) (Benchmark 3).

\begin{figure}
\centering
\begin{subfigure}{0.24\textwidth}
    \includegraphics[width=\textwidth]
    {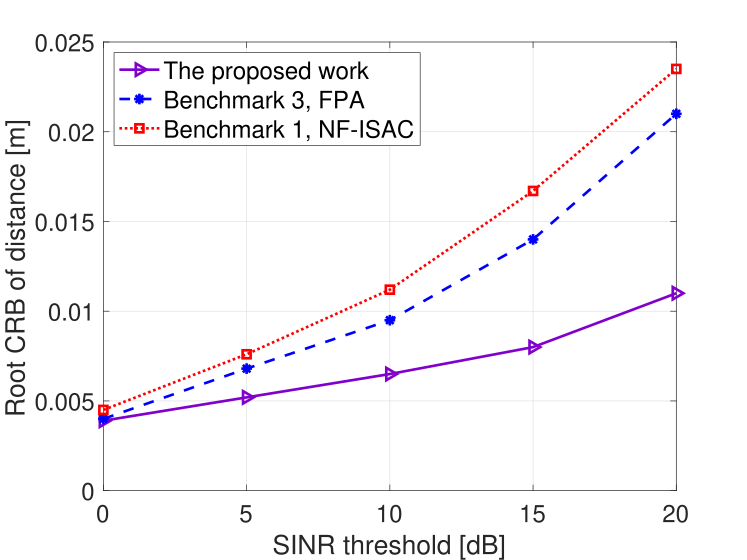}
    \caption{Distance}
    \label{SINR_distance}
\end{subfigure}
\hfill
\begin{subfigure}{0.24\textwidth}
    \includegraphics[width=\textwidth]{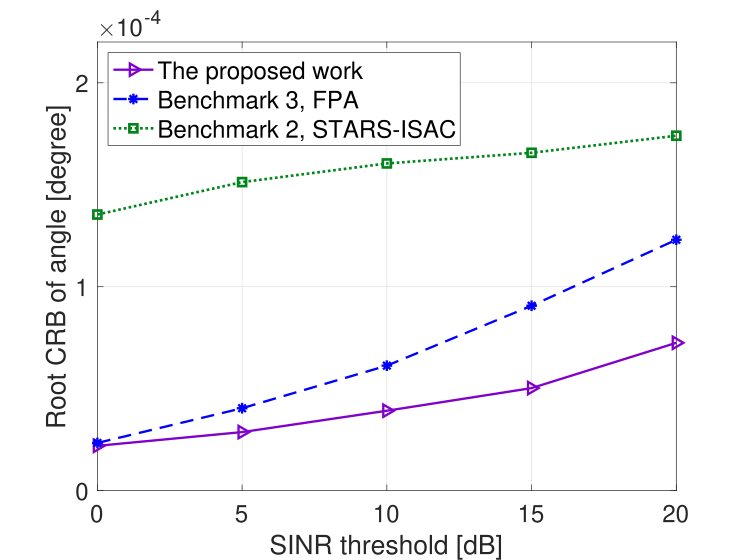}
    \caption{AoA}
    \label{SINR_angle}
\end{subfigure}
\caption{RCRB v.s. SINR threshold for $P$ = 40 dB and $K$ = 4}
\label{SINR}
\end{figure}

Fig. \ref{SINR} presents root CRBs (RCRBs) versus SINR threshold $\overline{\gamma}$ when $P = 40$ dB and $K = 4$. We can see that as $\overline{\gamma}$ increases, RCRBs tend to rise. This is due to the limited power budget. A greater portion of power is allocated to maintain the communication quality, thereby reducing the available power for sensing, and thus diminishing the sensing performance. Further, in the low $\overline{\gamma}$ regime, our RCRBs are close to RCRBs achieved by other benchmarks. This is led by the trade-off between communications and sensing in the optimization process. When $\overline{\gamma}$ is low, almost all resources and power budgets are allocated towards enhancing sensing capabilities, leading to nearly uniform performances for all algorithms. However, the performance gap between our work and other benchmarks is observed to expand as $\overline{\gamma}$ increases, demonstrating that our proposed algorithm is particularly effective when both communication and sensing requirements are stringent.

\begin{figure}
\centering
\begin{subfigure}{0.24\textwidth}
    \includegraphics[width=\textwidth]
    {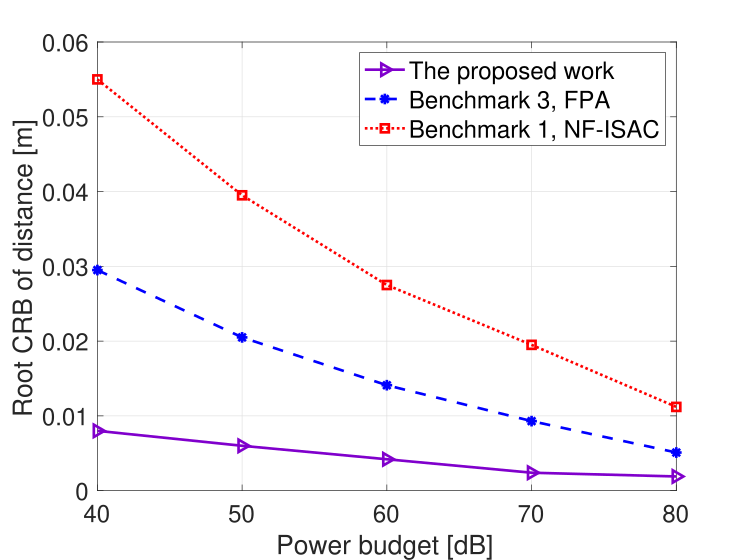}
    \caption{Distance}
    \label{power_distance}
\end{subfigure}
\hfill
\begin{subfigure}{0.24\textwidth}
    \includegraphics[width=\textwidth]{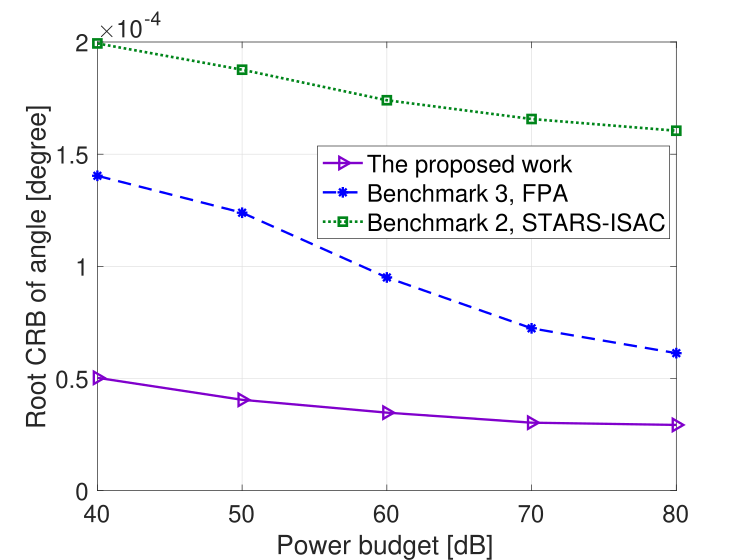}
    \caption{AoA}
    \label{power_angle}
\end{subfigure}
\caption{RCRB v.s. power budget for $\overline{\gamma}$ = 15 dB and $K$ = 4}
\label{power}
\end{figure}

Next, the relationship between power budget $P$ and RCRBs with $\overline{\gamma} = 15$ dB and $K = 4$ is shown in Fig. \ref{power}. We can see that as $P$ increases, more power becomes available to minimize the CRB, leading to an enhancement of the sensing performance. In addition, Fig. \ref{power_angle} demonstrates that compared with conventional FPA, the FA can significantly reduce the RCRB when $P = 80$ dB by providing more DoFs. Moreover, Fig. \ref{power_distance} shows the higher-dimensional sensing ability of the proposed near-field system, as the traditional far-field sensing can only obtain the AoA information of the target. 



\vspace{-0.4cm}


\section{Conclusion} 
This paper proposes a near-field ISAC system aided by the XL-STARS and the FA. To solve the formulated non-convex location optimization problem of the CRB, we introduce an iterative algorithm based on BCD and PDD methods. Simulation results demonstrate the superior performance of our system compared to other benchmarks. However, the FA movable region in our work is limited to one dimension. Thus, we leave the exploration of the near-field systems equipped with multidimensional FAs for future exploration.

\appendices

\section{Derivation of Fisher Information Matrices} \label{ap1}
According to \cite{kay1993fundamentals}, the FIM for vector $\xi$ from $\mathbf{y}_{s}$ is given by \begin{equation} \mathbf{J}_{\xi} = \frac{2\varsigma _{t}^{2}}{\sigma _{s}^{2}} \Re\left \{ \frac{\partial \mathbf{v}^{H}}{\partial_{\xi}} \frac{\partial \mathbf{v}}{\partial_{\xi}} \right \}.\end{equation}
\vspace{-0.35cm}

Since covariance matrix $\mathbf{R}_{x}\approx \frac{1}{T}\sum_{t=1}^{T} \mathbf{x}[t] \mathbf{x}^{H}[t]$, by defining $ \dot{\mathbf{G}}_q = \mathbf{a}(r_{q},\theta_{q}) \mathbf{a}^{T}(r_{q},\theta_{q})$, $\dot{\mathbf{G}}_{r_{q}} = \frac{\partial{\dot{\mathbf{G}}_q}}{\partial{r_q}}$, and $\dot{\mathbf{G}}_{\theta_{q}} = \frac{\partial{\dot{\mathbf{G}}_q}}{\partial{\theta_{q}}}$, we have $\frac{\partial \mathbf{v} }{\partial \theta _{q}} = \beta_q \text{vec}\left (\dot{\mathbf{G}}_{\theta_{q}} \mathbf{\Phi}_{r} \mathbf{H}_{br} \mathbf{X} \right )$, $\frac{\partial \mathbf{v} }{\partial r _{q}} = \beta_q \text{vec}\left (\dot{\mathbf{G}}_{r_{q}} \mathbf{\Phi}_{r} \mathbf{H}_{br} \mathbf{X} \right )$, and $\frac{\partial \mathbf{v} }{\partial \beta_q } = (1,j)\otimes \text{vec}\left (\dot{\mathbf{G}}_q \mathbf{\Phi}_{r} \mathbf{H}_{br} \mathbf{X} \right )$. Thus, for any $l, p \in \{r_q, \theta_q\}$, matrix $\mathbf{J}_{\xi}$ can be given by \vspace{-0.2cm}\begin{equation}
\begin{split}
J_{ll} = \frac{2\left | \beta_q  \right |^{2}T}{\sigma _{s}^{2}} \Re\left \{ \Tr\left ( \dot{\mathbf{G}}_{l}^{H} \mathbf{\Phi}_{r}^{H} \mathbf{H}_{br}^{H} \mathbf{R}_{x} \mathbf{H}_{br} \mathbf{\Phi}_{r} \dot{\mathbf{G}}_{l}\right ) \right \},   
\end{split}
\end{equation}
\vspace{-0.2cm}
\begin{equation}
\begin{split}
     J_{lp} 
     =  \frac{2\left | \beta_q  \right |^{2}T}{\sigma _{s}^{2}} \Re\left \{ \Tr\left ( \dot{\mathbf{G}}_{l}^{H} \mathbf{\Phi}_{r}^{H} \mathbf{H}_{br}^{H} \mathbf{R}_{x} \mathbf{H}_{br} \mathbf{\Phi}_{r} \dot{\mathbf{G}}_{p}\right ) \right \},
\end{split}
\end{equation}
\vspace{-0.2cm}
\begin{equation}
\begin{split}
     J_{l\beta_q^{r}} 
     =  \frac{2\beta_q^{*}T}{\sigma _{s}^{2}} \Re\left \{ \Tr\left ( \dot{\mathbf{G}}_l^{H} \mathbf{\Phi}_{r}^{H} \mathbf{H}_{br}^{H} \mathbf{R}_{x} \mathbf{H}_{br} \mathbf{\Phi}_{r} \dot{\mathbf{G}_q} \right ) \right \},
\end{split}
\end{equation}
\begin{equation}
\begin{split}
     J_{\beta_q^{r}\beta_q^{r}} 
     =  \frac{2T}{\sigma _{s}^{2}} \mathbf{I}_{2} \Tr\left ( \dot{\mathbf{G}}_q^{H} \mathbf{\Phi}_{r}^{H} \mathbf{H}_{br}^{H} \mathbf{R}_{x} \mathbf{H}_{br} \mathbf{\Phi}_{r} \dot{\mathbf{G}_q} \right ),
\end{split}
\end{equation}
\begin{equation}
\begin{split}
     J_{\beta_q^{i}\beta_q^{i}} = J_{\beta_q^{r}\beta_q^{r}},
     J_{\theta_{q}\beta_q^{i}}  = jJ_{\theta_{q}\beta_q^{r}}, J_{r_{q}\beta_q^{i}}  = jJ_{r_{q}\beta_q^{r}}.
\end{split}
\end{equation}

\section{Proof of the convergence of Algorithm 2}\label{ap2}

According to the BCD-based algorithm, we have \vspace{-0.2cm} \begin{equation}
\footnotesize
\begin{split}
    V^{(t-1)}_{obj} & = U(\mathbf{C}^{(t-1)}, \mathbf{O}^{(t-1)}, \mathbf{u}^{(t-1)}) \overset{(a)}{\geq} U(\mathbf{C}^{(t)}, \mathbf{O}^{(t-1)}, \mathbf{u}^{(t-1)}) \\
    & \overset{(b)}{\geq} U(\mathbf{C}^{(t)}, \mathbf{O}^{(t)}, \mathbf{u}^{(t-1)}) \overset{(c)}{=} U(\mathbf{C}^{(t)}, \mathbf{O}^{(t)}, \mathbf{u}^{(t)}) = V^{(t)}_{obj},
\end{split}
\end{equation} where (a) stems from that $\mathbf{C}^{(t)}$ is one suboptimal solution of \eqref{e4} with fixed coefficient vectors $\mathbf{O}^{(t-1)}$ of the XL-STARS and the APV $\mathbf{u}^{(t-1)}$, (b) is derived from that $\mathbf{O}^{(t)}$ is the optimal coefficient vector of \eqref{e4} with fixed $\mathbf{C}^{(t)}$ and $\mathbf{u}^{(t-1)}$, and (c) stems from that $\mathbf{u}^{(t)}$ is the optimal position vector of \eqref{e4} with fixed $\mathbf{O}^{(t)}$ and $\mathbf{C}^{(t)}$. Thus, \eqref{e4a} is non-decreasing when updating variables and is always positive, as it is finitely lower-bounded by zero; \textbf{Algorithm} \ref{a2} can always converge.

\bibliographystyle{ieeetr}
\bibliography{ref}

\vspace{12pt}
\color{red}

\end{document}